\documentclass[a4paper]{jpconf}
\usepackage{graphicx}
\begin{document}
\title{10 pc Scale Circumnuclear Molecular Gas Imaging of Nearby AGNs}

\author{Satoki Matsushita}

\address{Institute of Astronomy and Astrophysics, Academia Sinica,
	P.O.~Box 23-141, Taipei 10617, Taiwan, R.O.C.}
\address{Joint ALMA Observatory, Alonso de C\'ordova 3107,
	Vitacura 763 0355, Santiago, Chile}

\ead{satoki@asiaa.sinica.edu.tw}

\begin{abstract}
We present the images and kinematics of circumnuclear molecular gas
from 100 pc scale down to 10 pc scale in nearby active galactic
nuclei (AGNs) using the Submillimeter Array (SMA) and the Plateau de
Bure Interferometer (PdBI).
We have observed several nearby galaxies that host AGNs, such
as the nearest radio galaxy Centaurus A (NGC 5128), the Seyfert 2
galaxy M51 (NGC 5194), the Seyfert 2 galaxy NGC 1068, the Seyfert 1
galaxy NGC 1097, and the Seyfert 2 / starburst composite galaxy
NGC 4945, in CO lines to see whether the molecular gas distribution,
kinematics, and physical conditions at 10 -- 100 pc scale follows the
AGN unified model or not.
In 100 pc scale, most of the circumnuclear molecular gas shows
smooth velocity gradient, suggesting a regular rotating feature, and
also shows abnormal line ratios, suggesting the existence of active
sources to make the circumnuclear molecular gas dense and/or warm
conditions or abnormal chemical compositions.
In 10 pc scale, on the other hand, the molecular gas kinematics shows
various characteristics, some shows very disturbed kinematics such as
a jet-entrained feature in the galaxies that have jets, but some
still shows regular rotation feature in a galaxy that does not have
obvious jets.
These results indicate that the kinematics and physical/chemical
conditions of the circumnuclear molecular gas at the scale less than
100~pc is highly affected by the AGN activities, and at this scale,
there is no clear evidence of any unified feature seen in the
circumnuclear molecular gas.
\end{abstract}

\section{Introduction}
It was well known for a long time that there are two types of active
galactic nuclei (AGNs) at the centers of galaxies based on optical
spectra:
Type 1 AGNs show broad permitted lines as well as narrower forbidden
lines, and type 2 shows relatively narrower permitted and forbidden
lines \cite{wee77}.
In 1985, Antonucci and Miller \cite{ant85} published a result of a
polarization observation toward the nucleus of the type 2 AGN host
galaxy NGC 1068, showing that broad permitted lines have been
observed in the polarized (i.e., scattered) light, suggesting that
the type 1 AGN is hidden behind the dense obscuring material in this
type 2 nucleus.
This result has led to the AGN unification model, which is, the broad
line region is located close to a supermassive black hole of a galaxy
surrounded by a dense obscuring torus, and it will be type 1 if one
sees from the pole of the torus, and will be type 2 if on sees
through the torus \cite{ant93,urr95}.

\section{Circumnuclear Molecular Gas Distribution at 100 pc Scale}
After the suggestion of the AGN unification model, intense searches
of the obscuring torus have been started, especially using
millimeter-wave interferometers.
Jackson {\it et al} \cite{jac93} and Kohno {\it et al} \cite{koh96}
observed nearby Seyfert 2 galaxies NGC 1068 and M51 (NGC 5194),
respectively, with the HCN(1-0) line at a spatial resolution of
several arcseconds, and successfully imaged the dense gas around the
AGNs with the velocity gradients almost perpendicular to the radio
jets.
These results can be explained as dense gas disks or tori with the
radii of a few hundred pc, rotating around the AGNs perpendicular to
the radio jets.
Furthermore, the column density of the observed molecular gas of a
few $\times10^{23}$ cm$^{-2}$ \cite{koh96} was consistent with that
measured as the absorption column density in the X-ray observations
of $(4.2\pm1.5)\times10^{23}$ cm$^{-2}$ \cite{mak90}.
These results matched very well with the AGN unification model.

\begin{figure}[t]
\includegraphics[width=27pc]{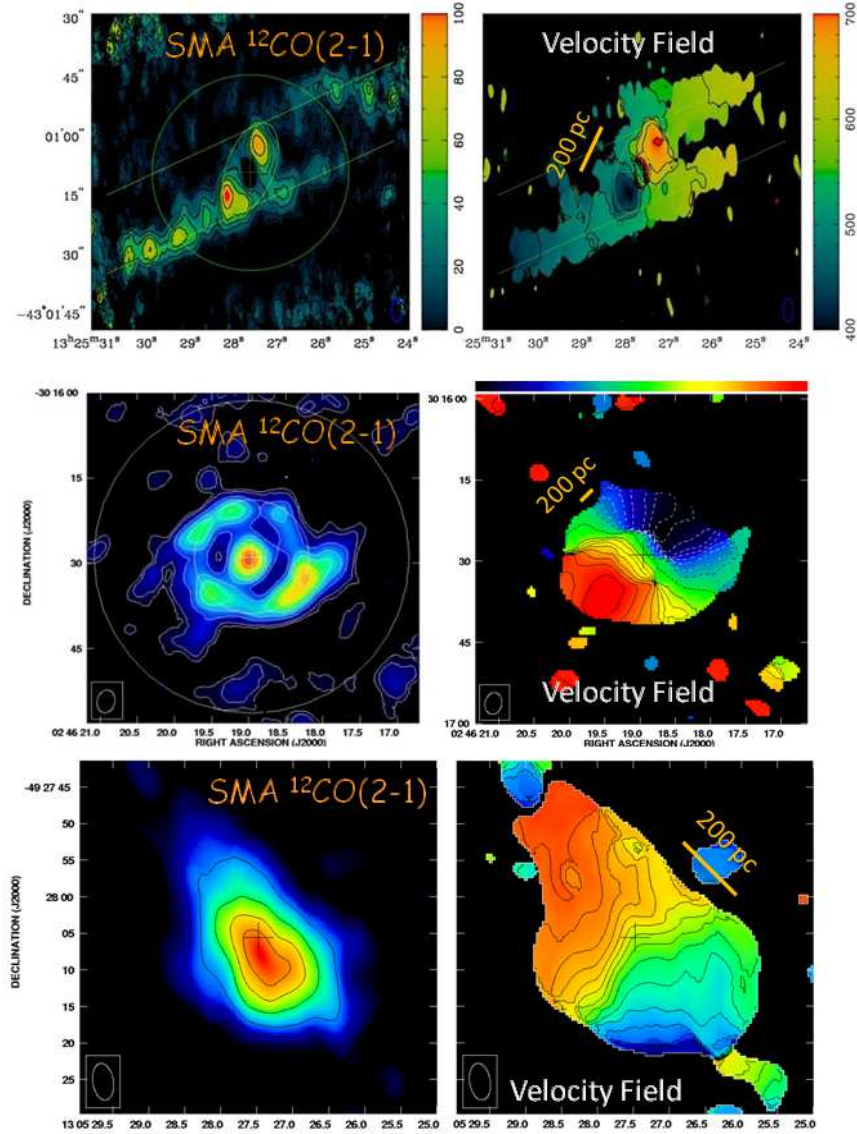}\hspace{0.5pc}
\begin{minipage}[b]{10pc}
\caption{\label{fig-100pc}
	CO(2-1) integrated intensity (moment 0; left) and
	intensity-weighted velocity field (moment 1; right) maps of
	{\it Above:} Centaurus A \cite{esp09}, {\it Middle:} NGC 1097
	\cite{hsi08}, and {\it Bottom:} NGC 4945 \cite{cho07}.
	All the images were obtained with the SMA.
	The linear scale of 200~pc is shown as a bar in the velocity
	field map of each galaxy.
	}
\end{minipage}
\end{figure}

Since the Submillimeter Array (SMA) \cite{ho04} has started
scientific observations, we have started to observe nearby AGNs in
the higher transition (J=2-1 and/or 3-2) CO lines at the spatial
resolutions of around a few hundred pc.
We observed the galactic centers (central a few kpc) of the Seyfert 1
galaxy NGC 1097 \cite{hsi08}, the Seyfert 2 / starburst composite
galaxy NGC 4945 \cite{cho07}, and the nearest radio galaxy Centaurus
A (NGC 5128) \cite{esp09} with the CO(2-1) line, and the obtained
images are shown in Figure~\ref{fig-100pc}.
All these galaxies have central molecular gas concentrations with the
scale of several hundred pc (note that Centaurus A has strong
absorption features at the galactic center, so that the image shows
a hole at the nucleus \cite{esp10}), and the velocity fields exhibit
smooth velocity gradient, namely rigid rotation feature, around the
nucleus (note that NGC 4945 is an edge-on galaxy with a bar, so that
the velocity field around the nucleus is affected by the non-circular
motion \cite{lin11}).
These velocity features again suggest rotating gas disks or tori with
the radii of a few hundred pc, similar as those observed in NGC 1068
and M51 mentioned above.
Since the kinematics of these galaxies shows the rigid rotation
feature, the gravitational fields of all these galaxies are dominated
by the bulge, not the supermassive black holes in the AGNs.

\begin{figure}[t]
\includegraphics[width=38pc]{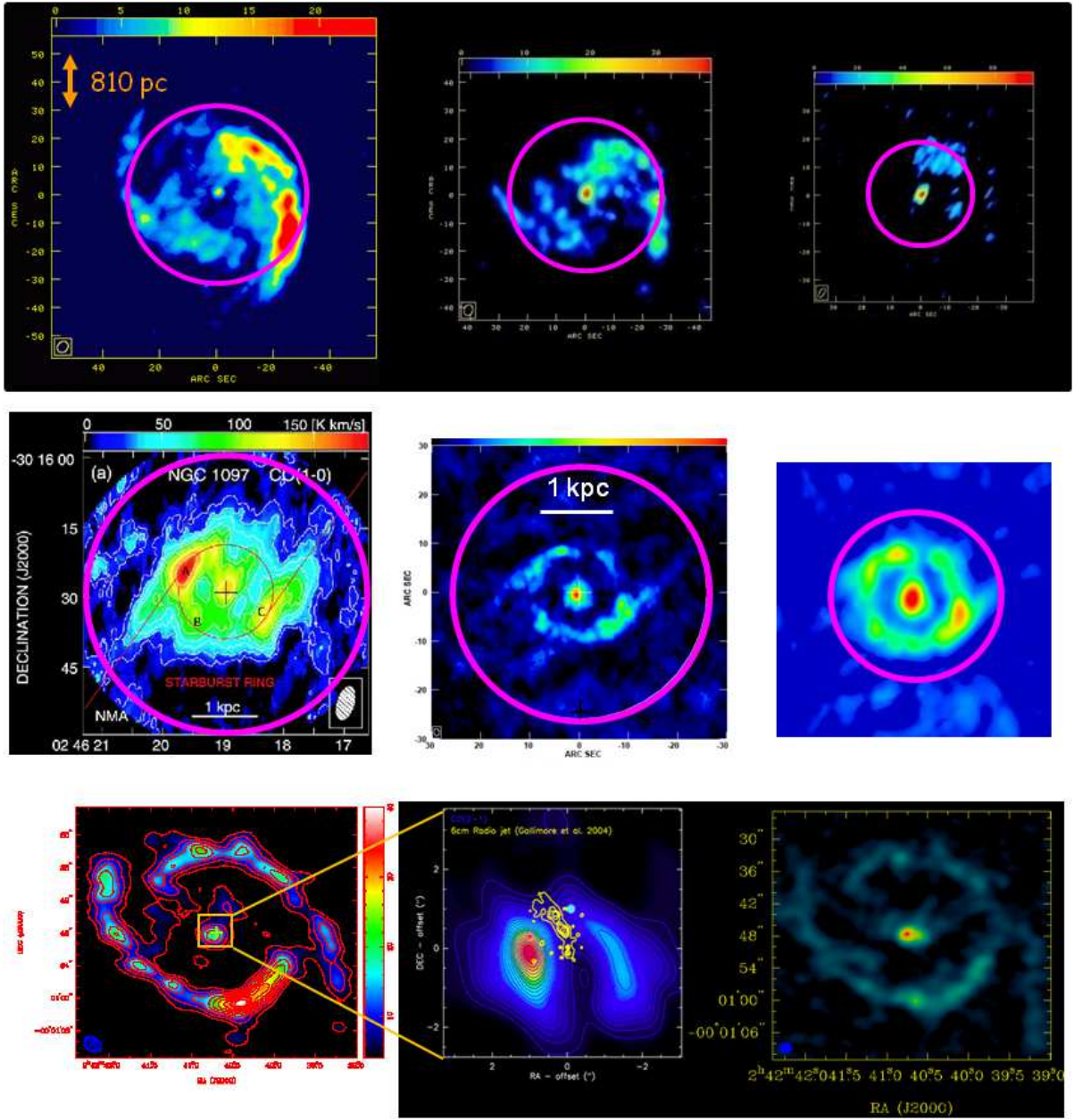}\hspace{0.5pc}
\begin{minipage}[b]{38pc}
\caption{\label{fig-multiline}
	Integrated intensity maps of {\it Left:} CO(1-0), {\it Middle:}
	CO(2-1), and {\it Right:} CO(3-2) toward {\it Top:} the Seyfert 2
	galaxy M51 (NGC 5194), {\it Middle:} the Seyfert 1 galaxy
	NGC 1097, and {\it Bottom:} the Seyfert 2 galaxy NGC 1068.
	Scales are matched for each galaxy, except for the CO(2-1) image
	of NGC 1068.
	The circles in the M51 and NGC 1097 images show the half power
	beam width of the primary antenna.
	M51 CO(1-0) is from Sakamoto {\it et al} (1999) \cite{sak99},
	M51 CO(3-2) from Matsushita {\it et al} (2004) \cite{mat04},
	NGC 1097 CO(1-0) from Kohno {\it et al} (2003) \cite{koh03},
	NGC 1097 CO(2-1) from Hsieh {\it et al} (2008) \cite{hsi08},
	NGC 1097 CO(3-2) from Hsieh {\it et al} (2011) \cite{hsi11},
	NGC 1068 CO(1-0) and CO(3-2) from Tsai {\it et al} (2012)
	\cite{tsa12}, and NGC 1068 CO(2-1) from Krips {\it et al} (2011)
	\cite{kri11} (in this image, the 6 cm radio jet contours are also
	overlaid \cite{gal04}).
	}
\end{minipage}
\end{figure}

On the other hand, we also obtained higher-J CO lines toward the
central a few kpc regions of NGC 1068 \cite{tsa12}, NGC 1097
\cite{hsi08,hsi11}, and M51 \cite{mat04}, and compared with the
previously observed lower-J CO line images
(Figure~\ref{fig-multiline}).
Central molecular condensations can be seen in all the images
(in whatever transitions) presented here, but the intensities are
stronger in higher-J lines than lower-J lines, indicating that
the circumnuclear molecular gas has either higher densities and/or
higher temperature conditions than, or abnormal chemical conditions
compared to the molecular gas at outer radii.
Lines ratios at outer radii, namely at the spiral arm regions or at
the starburst ring, have similar values as those regions in other
galaxies, so that this abnormal line ratios are obviously affected by
the existence of the AGNs.

These results suggest that the AGN activities {\it do not} affect the
kinematics of the circumnuclear molecular gas at 100 pc scale, but
{\it do} affect the physical/chemical conditions of the circumnuclear
molecular gas.

\section{Circumnuclear Molecular Gas Distribution at 10 pc Scale}
In our SMA M51 CO(3-2) data presented above, we noticed that the
kinematics of the circumnuclear molecular gas deviates from the
circular rotation as suggested before \cite{mat04}, but could not
study in detail due to the limited (large) spatial resolution.
We therefore observed the circumnuclear molecular gas of M51 with
$\sim10$~pc spatial resolution to spatially resolve the gas
kinematics around the Seyfert 2 nucleus in this galaxy.
We used the new-A (newly extended) configuration (maximum baseline
length $=760$~m) of the IRAM Plateau de Bure Interferometer (PdBI),
and observed with the CO(2-1) line, since the CO(2-1) line flux is
higher than that of the CO(1-0) line, and it is possible to obtain
higher spatial resolution images.

\begin{figure}[t]
\includegraphics[width=38pc]{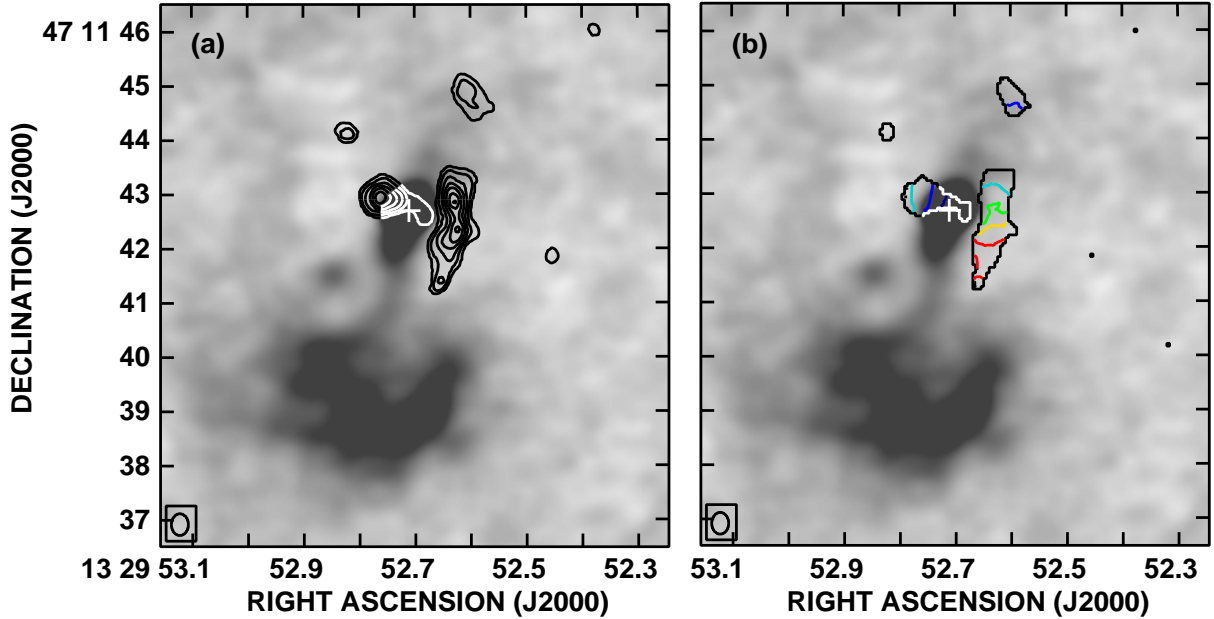}\hspace{0.5pc}
\begin{minipage}[b]{38pc}
\caption{\label{fig-co21}
	CO(2-1) {\it (a)} integrated intensity (moment 0) and {\it (b)}
	intensity-weighted velocity field (moment 1) maps of M51 at
	$0.41''\times0.31''$ (14~pc $\times$ 11~pc) resolution
	\cite{mat07} overlaid on the VLA 6~cm radio continuum greyscale
	image \cite{cra92}.
	The cross in each figure marks the position of the AGN.
	}
\end{minipage}
\end{figure}

Figure~\ref{fig-co21} shows the newly obtained circumnuclear CO(2-1)
image of M51 \cite{mat07}.
The spatial resolution reached to $0.41''\times0.31''$, which
corresponds to the linear scale of 14~pc $\times$ 11~pc at the
distance of M51 (7.1~Mpc \cite{tak06}).
Figure~\ref{fig-co21}(a) is the integrated intensity map of the
CO(2-1) data.
As you can see, there are mainly two molecular gas clouds at the
eastern and western sides of the nucleus, perpendicular to the radio
jets.
The cloud at the western side is elongated along north-south
direction, almost parallel to the radio jets, with the projected
distance to the nucleus of only $\sim30$~pc.
The one at the eastern side only has one strong peak with the size of
about 30~pc ($\sim1''$), roughly the size of the typical Giant
Molecular Cloud (GMC) in our Galaxy.
There is a weak emission extending toward the nucleus from this
cloud.
The column density of the molecular gas in front of the nucleus
calculated from this weak emission is very small, only
$6.2\times10^{21}$ cm$^{-2}$, much smaller than the column density of
atomic gas derived from the recent X-ray absorption measurements of
$5.6\times10^{24}$ cm$^{-2}$ \cite{fuk01}.
The missing flux due to our interferometric observations cannot
explain this difference; this is because the amount of the detected
flux has to be at the order of only 0.1\% of the total flux for
explaining the difference with the missing flux, but our observation
obviously detected much more than this flux level.
This suggests that the absorbing material in front of the X-ray
source (i.e., AGN) is much smaller than our beam size of $\sim10$ pc.

Figure~\ref{fig-co21}(b) is the intensity-weighted velocity field map
of the CO(2-1) data.
The eastern cloud exhibits the velocity gradient from west
(blueshift) to east (redshift).
This direction of the velocity gradient is totally opposite sense
with that observed in the past with $\sim10$ times larger spatial
resolution \cite{koh96}.
On the other hand, the western cloud exhibits the velocity gradient
from north (blueshift) to south (redshift), namely the velocity
gradient along the radio jets.
The amount of the velocity gradient of the western cloud is
$2.2\pm0.3$ km s$^{-1}$ pc$^{-1}$, which is consistent with that of
the ionized gas along the radio jets \cite{bra04}.
These distribution and the kinematics of the western cloud together
with the kinematics of the ionized gas along the radio jets suggest
that the western cloud is entrained by the radio jets \cite{mat07}.
Note that if we smooth our CO(2-1) data to ten times larger spatial
resolution, the kinematics is consistent with that of the past
observations \cite{koh96}, since the average velocities of the
eastern and the western clouds are consistent with the blueshifted
and the redshifted velocities, respectively, of the past
observations.

In summary, our $\sim10$~pc spatial resolution CO(2-1) results show
that there is no clear evidence of rotating disk or torus around the
Seyfert 2 nucleus of M51 as suggested in the past, but exhibit that
there are highly disturbed molecular gas features very close to the
nucleus, possibly affected by the AGN activities.
Furthermore, the obscuring material of this type 2 AGN has the size
much smaller than our beam size of $\sim10$ pc.

Such disturbed circumnuclear gas near AGNs is also recently suggested
in NGC 1068 \cite{kri11}:
Recent multiple molecular line observations toward the nucleus of
NGC 1068 at a spatial resolution of $\sim30$~pc with the SMA and the
PdBI revealed that the circumnuclear molecular gas kinematics is
better explained by a rotation with an outflowing motion, rather than
by a warped disk as suggested in the past \cite{sch00}.
The radius of the rotating disk is estimated to be around 60~pc.
The molecular outflow is suggested to be flowing along the radio jets
and the ionization cone with the velocity gradient of
3 km s$^{-1}$ pc$^{-1}$, similar to that observed in the
jet-entrained molecular gas in M51 (see above).

On the other hand, a circumnuclear molecular gas disk without any
disturbance has also been observed:
Recent CO(2-1) observation toward NGC 4945 at a spatial resolution of
$\sim20$~pc with the SMA has displayed a circumnuclear molecular gas
disk with no disturbance in the kinematics (i.e., rigid rotation
feature in the position-velocity diagram) \cite{lim11}.
The radius of the rotating disk is around 20~pc, which is closer to
the AGN than the distance of the circumnuclear clouds of M51 as
mentioned above.
Note that NGC 4945 does not have any obvious jet from its Seyfert 2
nucleus, in contrast to the nuclei of M51 and NGC 1068 that exhibit
radio and/or ionized gas jets.

\section{Discussion}
We only have three observations of the circumnuclear molecular gas in
$\sim10$~pc spatial resolution, but all the samples show different
features.
The circumnuclear gas of M51 does not show any evidence of clear
rotation feature, but show totally disturbed kinematics due to the
radio jets.
That of NGC 1068 exhibits rotation feature with outflowing gas along
the radio jets, but in the case of NGC 4945, which does not have
obvious jets from the AGN, only rotation feature can be seen.
These results suggest that the jet activity from the AGN affects the
circumnuclear molecular gas around $\sim10$~pc scale, which could not
be seen at $\sim100$~pc scale.
This suggests that there is no clear unified feature in the
circumnuclear molecular gas around AGNs at the scale of 10 pc.
Together with the line ratio anormaly, which means the physical
conditions or chemical composition anormaly, in the circumnuclear
molecular gas as mentioned above, the circumnuclear molecular gas
close to AGNs is suggested to be highly affected by the AGN
activities.

This also suggests that the star formation near AGNs can be highly
affected by the AGN activities; for instance, AGNs with strong jet
activities can have less star formation in the circumnuclear
(within 100~pc) regions, since too large turbulence induced by the
jet activities.
Due to very limited sample we have, it is difficult to say anything
about this kind of possibilities, but if we increase samples (with
ALMA for example), we may be possible to provide some hints for the
AGN-starburst connection.

\ack
We thank all the colleagues who support this SMA Seyfert Survey
program, including Jeremy Lim, Dinh-Van-Trung, Sebastien Muller,
Daniel Espada, Pei-Ying Hsieh, Richard C.-Y. Chou, Meng-Chun Tsai,
Ya-Lin Wu, Nagisa Oi, Kotaro Kohno, Kazushi Sakamoto, Frederic Boone,
and Melanie Krips.
This work is supported by the National Science Council (NSC) of
Taiwan, NSC 97-2112-M-001-021-MY3 and NSC 100-2112-M-001-006-MY3.
The Submillimeter Array is a joint project between the Smithsonian
Astrophysical Observatory and the Academia Sinica Institute of
Astronomy and Astrophysics and is funded by the Smithsonian
Institution and the Academia Sinica.


\section*{References}
\begin{thebibliography}{99}
\bibitem{wee77} Weedman D W 1977 {\it Ann. Rev. Astron. Astrophys.}
	{\bf 15} 69
\bibitem{ant85} Antonucci R R J and Miller J S 1985
	{\it Astrophys. J.} {\bf 297} 621
\bibitem{ant93} Antonucci R 1993 {\it Ann. Rev. Astron. Astrophys.}
	{\bf 31} 473
\bibitem{urr95} Urry C M and Padovani P 1995
	{\it Publ. Astron. Soc. Pacific} {\bf 107} 803
\bibitem{jac93} Jackson J M, Paglione T A D, Ishizuki S and
	Nguyen-Q-Rieu 1993 {\it Astrophys. J. Lett.} {\bf 418}, L13
\bibitem{koh96} Kohno K, Kawabe R, Tosaki T and Okumura S K 1996
	{\it Astrophys. J. Lett.} {\bf 461} L29
\bibitem{mak90} Makishima K, Ohashi T, Kondo H, Palumbo G G C and
	Trinchieri G 1990 {\it Astrophys. J.} {\bf 365} 159
\bibitem{ho04} Ho P T P, Moran J M and Lo F 2004
	{\it Astrophys. J. Lett.} {\bf 616} L1
\bibitem{hsi08} Hsieh P-Y, Matsushita S, Lim J, Kohno K and
	Sawada-Satoh S 2008 {\it Astrophys. J.} {\bf 683} 70
\bibitem{cho07} Chou R C Y, Peck A B, Lim J, Matsushita S, Muller S,
	Sawada-Satoh S, Dinh-V-Trung, Boone F and Henkel C 2007
	{\it Astrophys. J.} {\bf 670} 116
\bibitem{esp09} Espada D, {\it et al} 2009 {\it Astrophys. J.}
	{\bf 695} 116
\bibitem{esp10} Espada D, {\it et al} 2010 {\it Astrophys. J.}
	{\bf 720} 666
\bibitem{lin11} Lin, L-H, Taam R E, Yen D C C, Muller S and Lim J
	2011 {\it Astrophys. J.} {\bf 731} 15
\bibitem{tsa12} Tsai, M, Hwang C-Y, Matsushita S, Baker A J and
	Espada D 2012 {\it Astrophys. J.} {\bf 746} 129
\bibitem{hsi11} Hsieh P-Y, Matsushita S, Liu G, Ho P T P, Oi N and
	Wu Y-L 2011 {\it Astrophys. J.} {\bf 736} 129
\bibitem{mat04} Matsushita S, {\it et al} 2004 {\it Astrophys. J.
	Lett.} {\bf 616} L55
\bibitem{sak99} Sakamoto K, Okumura S K, Ishizuki S and
	Scoville N Z 1999 {\it Astrophys. J. Suppl.} {\bf 124} 403
\bibitem{koh03} Kohno K, Ishizuki S, Matsushita S, Vila-Vilar\'o B
	and Kawabe R 2003 {\it Publ. Astron. Soc. Japan} {\it 55} L1
\bibitem{kri11} Krips M, {\it et al} 2011 {\it Astrophys. J.}
	{\bf 736} 37
\bibitem{gal04} Gallimore J F, Baum S A and O'Dea C P 2004
	{\it Astrophys. J.} {\bf 613} 794
\bibitem{mat07} Matsushita S, Muller S and Lim J 2007 {\it Astron.
	Astrophys.} {\bf 468} L49
\bibitem{tak06} Tak\'ats K and Vink\'o J 2006, {\it Mon. Not. Royal
	Astron. Soc.} {\bf 372} 1735
\bibitem{cra92} Crane P C and van der Hulst J M 1992
	{\it Astron. J.} {\bf 103} 1146
\bibitem{fuk01} Fukazawa Y, Iyomoto N, Kubota A, Matsumoto Y and
	Makishima K 2001 {\it Astron. Astrophys.} {\bf 374} 73
\bibitem{bra04} Bradley L D, Kaiser M E and Baan W A 2004
	{\it Astrophys. J.} {\bf 603} 463
\bibitem{sch00} Schinnerer E, Eckart A, Tacconi L J, Genzel R and
	Downes D 2000 {\it Astrophys. J.} {\bf 533} 850
\bibitem{lim11} Lim J, Muller S, Boone F, Matsushita S and
	Dinh-V-Trung 2011 {\it Astrophys. J.} submitted
\end{thebibliography}
\end{document}